\documentclass{aa}  

\usepackage{graphicx}

\usepackage{txfonts}

\usepackage{epstopdf}
\usepackage{color}

\bibpunct{(}{)}{;}{a}{}{,}

\begin{document}

   \title{Spectrally-resolved UV photodesorption of CH$_4$ in pure and layered ices} 

   \author{R. Dupuy\inst{1},
		M. Bertin\inst{1}, 
		G. F\'{e}raud\inst{1},
		X. Michaut\inst{1},
		P. Jeseck\inst{1},
		M. Doronin\inst{1},
		L. Philippe\inst{1},
		C. Romanzin\inst{2}
          \and
          J.-H. Fillion\inst{1}
          }

   \institute{Laboratoire d'Etude du Rayonnement et de la Matière en Astrophysique et Atmosphères (LERMA), Sorbonne Universités, UPMC Univ. Paris 06, Observatoire de Paris, PSL Research University, CNRS UMR 8112,  F-75005, Paris, France\\
        \and
            Laboratoire de Chimie Physique (LCP), CNRS UMR 8000, Univ. Paris Sud, F-91400 Orsay, France}

   \date{}

 
  \abstract
   {Methane is among the main components of the ice mantles of insterstellar dust grains, where it is at the start of a rich solid-phase chemical network. Quantification of the photon-induced desorption yield of these frozen molecules and understanding of the underlying processes is necessary to accurately model the observations and the chemical evolution of various regions of the interstellar medium.}
   {This study aims at experimentally determining absolute photodesorption yields for the CH$_4$ molecule as a function of photon energy. The influence of the ice composition is also investigated. By studying the methane desorption from layered CH$_4$:CO ice, indirect desorption processes triggered by the excitation of the CO molecules is monitored and quantified.}
   {Tunable monochromatic VUV light from the DESIRS beamline of the SOLEIL synchrotron is used in the 7 - 13.6 eV (177 - 91~nm) range to irradiate pure CH$_4$ or layers of CH$_4$ deposited on top of CO ice samples. The release of species in the gas phase is monitored by quadrupole mass spectrometry and absolute photodesorption yields of intact CH$_4$ are deduced.}
   {CH$_4$ photodesorbs for photon energies higher than \textasciitilde9.1 eV (\textasciitilde136 nm). The photodesorption spectrum follows the absorption spectrum of CH$_4$, which confirms a desorption mechanism mediated by electronic transitions in the ice. When it is deposited on top of CO, CH$_4$ desorbs between 8 and 9 eV with a pattern characteristic of CO absorption, indicating desorption induced by energy transfer from CO molecules.}
   {The photodesorption of CH$_4$ from the pure ice in various interstellar environments is around $2.0 \pm 1.0 \times 10^{-3}$ molecules per incident photon. Results on CO-induced indirect desorption of CH$_4$ provide useful insights for the generalization of this process to other molecules co-existing with CO in ice mantles.}

    \titlerunning{UV Photodesorption of CH$_4$}
    \authorrunning{R. Dupuy et al.}

   \maketitle

\section{Introduction}
In cold and dense regions of the interstellar medium (ISM), such as dense molecular clouds or protoplanetary disks midplanes, interstellar dust grains are coated with an ice mantle, composed of molecules either accreted from the gas phase or formed directly on the surface. Due to the low temperatures in these regions ($\sim$ 10 K) thermal desorption is negligible, and the exchanges between the gas phase and those icy mantles are thus driven by non-thermal processes. Among them, desorption stimulated by UV photons emitted by nearby stars or secondary UV photons emitted by excited molecules is thought to play a significant role. Laboratory experiments in the last few years have aimed at quantitatively and qualitatively describing such photon-stimulated desorption (PSD) processes for astrophysically relevant ice analogs, \citep[see e.g.][]{oberg_photodesorption_2009,hama_translational_2009,fayolle_co_2011,yuan_radiation_2014,martin-domenech_study_2016} which is required in order to explain the observations of molecule abundances in the gas and solid phase and to accurately model the chemical evolution in the concerned regions of the ISM.
Photodesorption can also be relevant in other astrophysical media: it has been considered as one of the possible space-weathering agents of solar system bodies \citep{bennett_space-weathering_2013}.

CH$_4$ has been observed towards protostellar objects \citep{dartois_molecular_1998,boogert_gas/solid_1998} in the gas and solid phase, with an estimated CH$_4$ gas to ice ratio of 0.3-0.5. CH$_4$ formation was proposed to occur on grains, and it was found to exist in polar ices containing H$_2$O and CH$_3$OH \citep{boogert_gas/solid_1998}.
CH$_4$ is also detected in ices towards young stellar objects (YSOs) with a typical abundance of 1-10~\% with respect to H$_2$O, while in quiescent clouds and cores only an upper limit of <3~\% is found \citep{boogert_observations_2015}. In cometary ices, the typical abundance relative to H$_2$O is 1~\% \citep{mumma_chemical_2011}, lower than in the previously cited environments. The observations toward YSOs suggest that the dominant formation mechanism of CH$_4$ in the ISM is through hydrogenation of atomic C at the surface of grains \citep{oberg_spitzer_2008}. Such a mechanism has been experimentally confirmed \citep{hiraoka_gasgrain_1998}. This implies that desorption from grains constitutes the main source of CH$_4$ in the gas phase. CH$_4$ abundances in the gas phase are poorly known mainly because of its lack of a permanent dipole moment, meaning that no detection is possible at radio wavelengths \citep{boogert_methane_2004}. This lack of observational constraints makes it even more crucial to have accurate modeling parameters when assessing the available reservoir of CH$_4$. Indeed, methane is included in gas-grain models for cold clouds, disks and illuminated regions such as photon-dominated regions (PDR) \citep{ruaud_gas_2016,kalvans_effect_2015,hollenbach_water_2009,garrod_formation_2011,cleeves_exploring_2016}.

Evaporation of CH$_4$ from the grain mantles is the first step of a gas phase chemistry leading to the formation of carbon chains (as in protostars and protoplanetary disks, see e.g. \cite{sakai_warm_2013,aikawa_physical_2006,hassel_modeling_2008}). Moreover, photodissociation products of methane initiate a chemical network that has been well experimentally studied in the gas phase, due to the fact that it is also very relevant to planetary science \citep{romanzin_methane_2005,gans_photolysis_2011,smith_modeling_1999}.

In the solid phase, CH$_4$ is also at the start of a rich chemical network. Laboratory experiments on the processing of pure CH$_4$ ice with energetic particles \citep{kaiser_theoretical_1998, baratta_ion_2003, bennett_laboratory_2006, de_barros_cosmic_2011} and UV photons \citep{lo_far-ultraviolet_2015, bossa_methane_2015} as well as mixes with other astrophysically relevant molecules such as H$_2$O \citep{oberg_effect_2010, hodyss_photochemistry_2009}, N$_2$ \citep{wu_spectra_2012}, CO \citep{maity_formation_2014} or N$_2$:CO \citep{hodyss_formation_2011} have yielded a great variety of products. 

Non-thermal desorption/sputtering of methane by energetic ions and electrons has already been extensively studied \citep[][and references therein]{gudipati_sputtering_2013}. This is not the case for UV irradiation. \citet{cruz-diaz_negligible_2016} observed CH$_4$ desorption when UV-irradiating a pure methanol ice, demonstrating another pathway to the release of methane in the gas phase than direct methane photodesorption. However, in their study of pure CH$_4$ ice, they did not observe intact CH$_4$ photodesorption \citep{cruz-diaz2015}. Instead, desorption of photoproducts such as C$_2$H$_4$ was seen. A recent study by \cite{martin-domenech_study_2016} has also shown that a part of the reaction products (e.g. H$_2$CO) initiated by the methane photodissociation in mixed CH$_4$:H$_2$O ices is released in the gas phase during UV irradiation. However, they did not observe intact CH$_4$ photodesorption either. The present study is therefore, to our knowledge, the first report of CH$_4$ photodesorption from pure CH$_4$ ice. We have focused on the desorption of the intact CH$_4$ molecule, and obtained absolute photodesorption yields as a function of the UV photon energy between 7 and 13.6 eV. This allows to derive photodesorption yields for any given astrophysically relevant UV field. In order to test the influence of the environment of the molecules on photodesorption, we also investigated the desorption of layers of CH$_4$ deposited on top of CO ice. CO is a very abundant component of interstellar ices, and has a high photodesorption yield \citep{fayolle_co_2011}, therefore its ability to indirectly induce desorption of other molecules \citep{bertin_indirect_2013} could have important astrophysical implications.

\section{Methods}

Experiments are performed in the SPICES (Surface Processes \& ICES) set-up, which has been described in detail previously \citep{doronin_adsorption_2015}. It consists of an ultra-high vacuum (UHV) chamber with a base pressure of typically $10^{-10}$ mBar, within which a polycrystalline gold surface is mounted on a rotatable cold head that can be cooled down to \textasciitilde 10~K using a closed cycle helium cryostat. Ices of $^{13}$CH$_4$ (Eurisotop, >99.9\% purity in $^{13}$C) and CO (Air liquide, >99.9\% purity) are dosed by exposing the cold surface (10~K) to a partial pressure of gas using a tube positioned a few millimeters away from the surface, allowing rapid growth without increasing the chamber pressure to more than a few 10$^{-9}$ mBar. Ice thicknesses are controlled with a precision better than 1 monolayer (ML) via a calibration using the temperature programmed desorption (TPD) technique, as detailed in \citet{doronin_adsorption_2015}. The release of species in the gas phase is monitored by means of a quadrupole mass spectrometer (QMS).

The chamber is then coupled to the undulator-based DESIRS beamline \citep{nahon_desirs:_2012} at the SOLEIL synchrotron facility, which provides a monochromatic, tunable radiation beam hitting the sample at 45$^{\circ}$ incidence. The coupling is window-free to prevent cut-off of the higher energy photons. To acquire photodesorption spectra, the narrow bandwith (\textasciitilde25 meV) output of a grating monochromator is continuously scanned between 7 and 14 eV. Higher harmonics of the undulator are suppressed using a Krypton gas filter. Typical photon fluxes as measured with a calibrated AXUV photodiode depend on the photon energy and vary between $5 \times 10^{12}$ and $17 \times 10^{12}$ photons.cm$^{-2}$.s$^{-1}$.

During the photon energy scan, the relative amount of photodesorbed molecules is recorded by the QMS. Each 25 meV photon energy step lasts about 5 s, which is sufficiently higher than the dwell time of the QMS (0.5 s). A typical scan thus lasts around 20 minutes, which corresponds to a total photon dose of \textasciitilde 10$^{16}$ photons.cm$^{-2}$. 

Once corrected from the background contribution and divided by the photon flux, the QMS signal I$_i$ corresponds to the relative efficiency of photodesorption as a function of photon energy. This signal can then be converted to absolute photodesorption yields. For a given species $i$, the absolute photodesorption yield Y$_i$ (in molecules per incident photons) is proportional to the signal $I_i$ at the m/z ratio of species $i$, such that $Y_i = f_i \times I_i$. $f_i$ can be further expanded into $f_i = g \times h_i$ where g is a factor intrinsic to the QMS and the geometry of our set-up, and $h_i$ is a factor specific to each species $i$. In the case of CO, $f_{CO}$ can be obtained by comparing the photodesorption signal of a pure CO ice taken during the experimental runs to the well-known CO absolute photodesorption yields \citep{fayolle_co_2011}. This calibration is then used as a basis to derive the proportionality coefficients $f_i$ of the other species. This is done by correcting $f_{CO}$ by molecule-dependent factors such as the electron-impact ionization cross-section, which characterizes the fact that CO and species $i$ are not ionized with the same efficiency by the ionization chamber, and the apparatus function of our QMS, which characterizes the fact that ions with a different m/z ratio are not detected with the same efficiency by the QMS. Thus, for CH$_4$, we have: $$Y_{CH_4} = f_{CH_4} \times I_{CH_4} = f_{CO} \times \frac{\sigma(CH_4^+/CH_4)}{\sigma(CO^+/CO)} \times \frac{AF(CH_4)}{AF(CO)} \times  I_{CH_4} $$
Where $\sigma(i^+/i)$ is the partial electron-impact ionization cross-section of the i$^+$ ion relatively to its parent molecule $i$ and $AF(i)$ is the apparatus function of our QMS for species $i$ ($AF(CO)$ is set to 1). 

So far we assumed that the QMS signal at a given m/z ratio originated from a single species $i$. However, due to cracking in the QMS, an ion can originate from different parent molecules, and the signal has to be corrected from the contribution of the irrelevant parent molecules. For example in our case, the signal measured on mass 16 (corresponding here to $^{13}$CH$_3^+$) could be originating from either desorbed CH$_3$ radicals or from desorbed CH$_4$ molecules that are cracked in the ionization chamber of the QMS. Thus in order to know if there is a signal originating from desorbing CH$_3$, the latter contribution has to be substracted using the partial ionization cross-sections of CH$_4$ and CH$_3$: $$ I_{CH_3} = I_{CH_3^+} - \frac{\sigma(CH_3^+/CH_4)}{\sigma(CH_4^+/CH_4)} \times \frac{AF(CH_3)}{AF(CH_4)} \times  I_{CH_4} $$
Where I$_{CH_3^+}$ is the total signal measured on mass channel 16 and  I$_{CH_3}$ is the signal originating from desorbing CH$_3$ (so far we had assumed that ions in the QMS had only one parent molecule, i.e. I$_i$ = I$_{i^+}$).

For similar reasons, $^{13}$CH$_4$ was used in order to distinguish it in mass spectrometry from O$^+$ fragments (m/z = 16, same as $^{12}$CH$_4^+$) originating from the cracking of CO in the QMS, during CH$_4$ layered on CO experiments. This consideration also brings up the possibility that part of the CH$_4^+$ signal detected originates not from desorbing CH$_4$ but from larger molecules formed by photolysis of the ice and photodesorbed. However, the expected photolysis products of solid CH$_4$, such as C$_2$H$_2$, C$_2$H$_4$, C$_2$H$_6$ or larger hydrogenated carbon-chain (e.g. \cite{lo_far-ultraviolet_2015}) only lead to negligible CH$_4^+$ signal by electron-impact ionization at 70 eV.\footnote{http://webbook.nist.gov/chemistry/} We therefore have made the assumption that desorption of possible photoproducts from the solid methane do not participate to the recorded CH$_4^+$ signal, that is thus only related to the desorption of intact CH$_4$.

The calibration of our spectra is done using the apparatus function of our QMS as determined by calibration experiments using several molecules with well-known electron-impact cracking patterns and the partial electron-impact ionization cross-sections at 70 eV found in \citet{tian_cross_1998} for CH$_4$ and CO. While the uncertainties on the relative photodesorption spectra are only due to experimental noise, background substraction and photon flux uncertainties, this calibration method introduces much larger uncertainties on the absolute photodesorption yields. We estimate a systematic 50\% uncertainty on all the absolute photodesorption yields given for CH$_4$ hereafter. 

\section{Results}

\subsection{Pure CH$_4$ ice}

\begin{figure}
    \resizebox{\hsize}{!}{\includegraphics{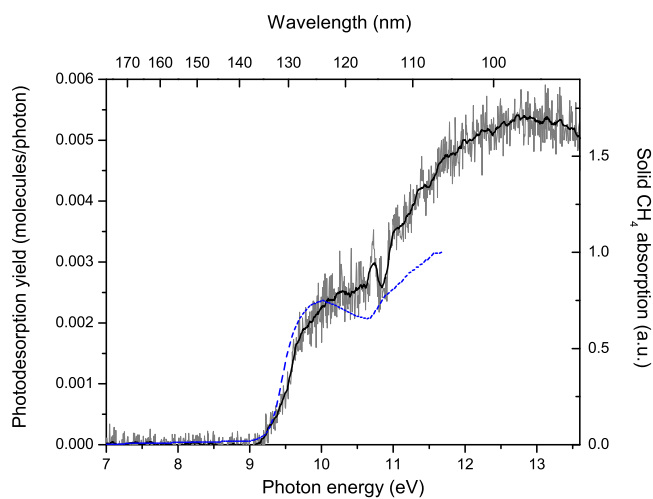}}
    \caption{$^{13}$CH$_4$ photodesorption spectrum of a 20 ML thick $^{13}$CH$_4$ ice at 10~K between 7 and 13.6 eV. The blue dotted line (right axis) is the absorption spectrum of pure CH$_4$ adapted from \citet{wu_spectra_2012}.}
    \label{fig1}
\end{figure}

Figure\ref{fig1} shows the photodesorption yield as a function of photon energy of CH$_4$ from a pure, 20 ML-thick CH$_4$ ice. There is a threshold at 9.1 eV under which desorption is below our detection limit. A broad peak is centered around 10 eV, then desorption steadily increases from 10.5 eV onwards, with a maximum around 13 eV. The absorption spectrum of pure CH$_4$ adapted from \citet{wu_spectra_2012} is given for comparison, and it matches fairly well the desorption profile, at least until their cut-off energy. The 10 eV broad feature is less well defined in our desorption spectrum. 
The absorption spectrum taken by \citet{koch_optical_1971}, which has a higher energy cut-off, suggests that the absorption keeps increasing until around 14 eV. We did not use that spectrum here because it has an 0.5 eV red-shift compared with the spectrum from \citet{wu_spectra_2012}. This is probably because of different experimental conditions: the spectrum by \citet{koch_optical_1971} was taken at 30K, where CH$_4$ is expected to be crystalline rather than amorphous \citep{hudson_activation_2015}. The spectrum from \citet{cruz-diaz_vacuum-uv_2014-1}, taken at 10K under UHV conditions as well and recorded at energies lower than 10.3~eV, agrees with the one by \citet{wu_spectra_2012}.

The absorption spectrum of CH$_4$ in the gas phase has a threshold at 8.5 eV and two, rather than one, broad features centered at 9.7 and 10.5 eV. These two features have been attributed to the 1t$_2$ -> 3s Rydberg transitions of respectively the D$_{2d}$ and C$_{3v}$ symmetries of the ion \citep{au_valence_1993}. As discussed in \citet{robin_higher_1974}, the equivalent feature at 10 eV in the solid spectrum can thus probably be attributed to excitons, which is also supported by calculations \citep{kunz_electronic_1983}. The fact that the desorption spectrum matches the absorption spectrum and the attributions discussed here is a clear evidence that desorption is initiated by an electronic transition. After 11 eV, the absorption of gas-phase CH$_4$ keeps increasing as in the solid phase, although there are no clearly discernible features. Virtual valence orbitals are thought to be above the first ionization potential at 13.6 eV, therefore the absorption in this region has been attributed to overlapping Rydberg transitions. The ionization threshold of CH$_4$ in the gas phase is at 12.5 eV \citep{kameta_photoabsorption_2002} and is red-shifted by 1.2 eV in the solid phase \citep{himpsel_ultraviolet_1975} which puts it at 11.3 eV. In the gas phase, \citet{kameta_photoabsorption_2002} have measured the relative weight of neutral dissociation and ionization in the absorption of CH$_4$. Below the ionization threshold, there is only neutral dissociation, but as the ionization channel opens, neutral dissociation decreases until only ionization processes occur, at 16 eV. Therefore, the desorption mechanisms at the origin of the 10 eV feature are likely different from those responsible for the desorption signal above 11 eV.   

We have observed evidence of aging of the CH$_4$ ice: when it has already been processed by VUV irradiation, the desorption yield decreases. This means that despite our low fluence, which usually allows to probe desorption processes without affecting the ice significantly, photochemistry and/or ice restructuration (due to release of energy in the bulk) must play a role. Systematic studies of the effect of processing will be required to understand the exact nature of the aging effect and how much it affects photodesorption. In the meantime we only present results for a "fresh" ice.

The masses corresponding to the CH$_3$, CH$_2$ and CH fragments were also monitored during the photon irradiation. Once corrected from the cracking of desorbing CH$_4$ in the ionization chamber of the QMS, the remaining signals fall below our sensitivity level, meaning that we are unable to observe the desorption of these radicals if they do desorb. This is despite the fact that at least the CH$_3$ radical fragment has been detected as trapped in photolyzed methane ice \citep{lo_far-ultraviolet_2015} and matrix-isolated methane \citep{milligan_infrared_1967}.

\subsection{CH$_4$ on top of CO ice}

\begin{figure}
    \resizebox{\hsize}{!}{\includegraphics{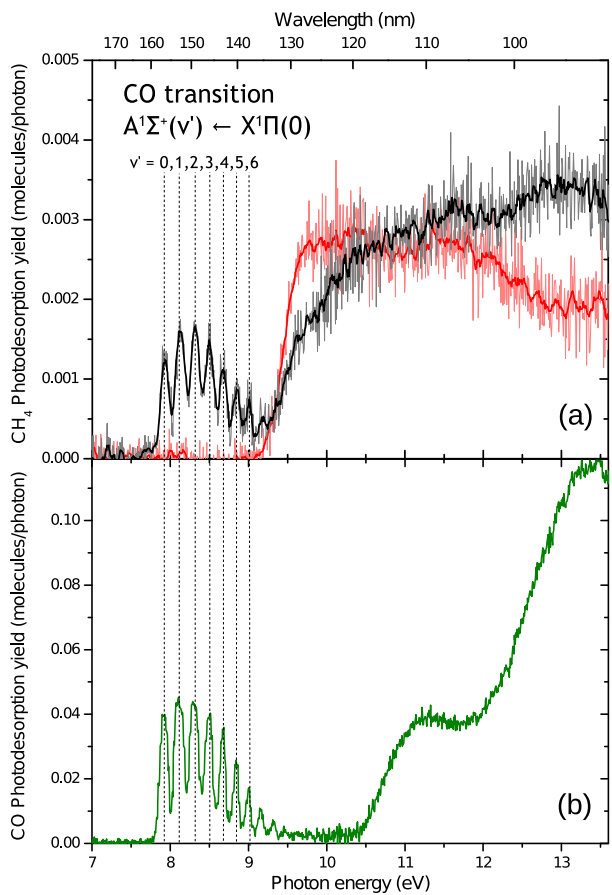}}
    \caption{a. $^{13}$CH$_4$ photodesorption spectrum of 1 ML (black trace) and 3 ML (red trace) of $^{13}$CH$_4$ on top of 20 ML of CO ice at 10~K, between 7 and 13.6 eV. 
    b. CO photodesorption spectrum of a 20 ML thick pure CO ice at 10~K.  
    Electronic transitions are indicated by vertical dashed lines.
    }
    \label{fig2}
\end{figure}

Figure~\ref{fig2}b shows the photodesorption spectrum of a pure 20 ML-thick CO ice (measured at m/z=28 CO). Between 8 and 9 eV, this spectrum shows very characteristic peaks that correspond to the vibrational levels of the A$^1\Sigma \leftarrow$ X$^1\Pi$ electronic transition of condensed CO. This feature allowed to label the desorption at these energies as a DIET (Desorption Induced by Electronic Transition) process \citep{fayolle_co_2011}. It was later found that desorption in this region could be in great part attributed to an indirect mechanism \citep{bertin_uv_2012}. "Indirect" here means that the desorbing molecule is not the one that has absorbed a photon. Instead, a CO molecule in the first few sub-surface monolayers of the ice is excited, then an energy transfer induces the desorption of another, surface-located CO molecule. Above 10.5~eV, two broad features are observed. The CO desorption mechanism for these two features is not clear, since in this region electronic states are dissociative and ionization can occur at 12.5 eV and above, which opens many more possibilities other than a "simple" DIET mechanism, such as photo-electron induced desorption or neutral-neutral or ion-neutral recombination chemistry.

\begin{figure}
    \resizebox{\hsize}{!}{\includegraphics{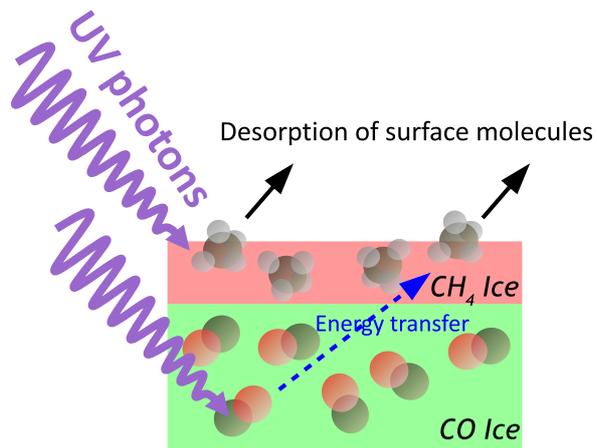}}
    \caption{Simplified scheme of the two desorption mechanisms at play: direct desorption of surface $^{13}$CH$_4$, and indirect desorption of surface $^{13}$CH$_4$ after energy transfer from underlying CO molecules.
    }
    \label{fig3}
\end{figure}

Figure~\ref{fig2}a shows the photodesorption spectrum of 1 ML of CH$_4$ deposited on top of 20 ML of CO. The desorption profile is significantly changed as compared with the pure CH$_4$ ice. Between 8 and 9 eV, where CH$_4$ does not desorb in the pure ice case (Fig.~\ref{fig1}), desorption peaks are observed that correspond to the signature peaks of CO excitation. This is a clear indication that an indirect desorption mechanism is at play. This indirect mechanism is schematized in Figure~\ref{fig3}: the excitation of a sub-surface CO molecule is followed by an energy transfer to a CH$_4$ molecule at the surface, which then desorbs. The principle is identical to the indirect mechanism that was found in pure CO, and we discuss in section 4.2 the generalization of this "CO-induced" desorption mechanism.
The photodesorption of CH$_4$ on CO is a superimposition of the pure CO (Fig.~\ref{fig2}b) and pure CH$_4$ (Fig.~\ref{fig1}) profiles: while the peaks between 8 and 9 eV are entirely due to CO-induced indirect desorption, desorption around 10 eV unambiguously corresponds to CH$_4$ absorption and subsequent desorption, since CO absorption (and desorption) is very weak in this region. In the region above 11 eV, both molecules absorb significantly and they probably both contribute to the observed desorption of CH$_4$, although no clear features can be discerned. 

On Figure~\ref{fig2}a is also represented the photodesorption spectrum of 3 ML of CH$_4$ deposited on top of 20 ML of CO. At this thickness, no CO-induced desorption is seen at all, and the spectrum looks like the photodesorption spectrum of 20 ML of CH$_4$ below 10.5~eV. In this system, indirect desorption is negligible if more than 3~ML of CH$_4$ are deposited on top of CO ice. Both the 1 and 3 ML layers of CH$_4$ on CO exhibit lower desorption yields in the region above 11 eV than in the case of the pure 20 ML ice, which suggests either a thickness dependence of the desorption in this region or that the ageing effect mentioned earlier is more efficient in thin layers, in which case the ice could have aged as the scan occurred, resulting in a lower yield at high energies. The yield at high energy of the 3 ML layer is even lower than the 1 ML one, which is evidence that there is a contribution, although it is almost structureless, of CO-induced desorption in the latter case.

\section{Discussion} 

\subsection{Photodesorption mechanisms in pure CH$_4$ ices}

In the relatively simpler case of molecules that do not dissociate upon VUV irradiation (e.g. CO under 10 eV, N$_2$), photodesorption has been attributed to a DIET process, which as discussed above can be direct or indirect, and where the molecule stays intact throughout the process \citep{bertin_uv_2012, bertin_indirect_2013}. Such a process cannot be excluded here, but  the fact that the energy threshold for dissociation of CH$_4$ in the gas phase is much lower than the photon energy \citep{gans_photolysis_2011} suggests that new desorption pathways should  be opened by photodissociation. Photodesorption initiated by molecular photodissociation is known to dominate in the case of H$_2$O. Molecular dynamics simulations on the photodesorption of H$_2$O ice \citep{andersson_photodesorption_2008,andersson_theoretical_2011} have suggested a kick-out mechanism by energetic H fragments. This mechanism could be generalized to other molecules such as CH$_4$ that produce energetic fragments, including H, when photodissociated. Another possible process is where desorption occurs as the fragments of a molecule that dissociated immediately recombine (geminate or direct recombination). Recombination can also occur for fragments coming from different molecules (secondary recombination): the fact that fragments are very energetic means that in condensed phase, they can travel far on the surface and in the bulk and react with other fragments. Such a process has been shown to exist for the UV photodesorption of O$_2$ \citep{fayolle_wavelength-dependent_2013} and CO$_2$ \citep{fillion_wavelength_2014, martin-domenech_uv_2015}. Another indication that photodissociation and subsequent radical chemistry likely plays an important role in CH$_4$ desorption, is the desorption of heavier hydrocarbons such as C$_2$H$_4$ observed by \citet{cruz-diaz2015} upon VUV irradiation of CH$_4$ ice. 

One interesting piece of information that we can extract from our data is where the energy goes when photons are absorbed at the surface of the ice. The final outcomes of the initial event of a photon being absorbed by a molecule near the surface can be divided into three channels. Two measurable channels are (i) a CH$_4$ desorption event and (ii) the effective destruction of a CH$_4$ molecule. It should be noted that the second channel does not characterize photodissociation, since photodissociation can also lead to desorption or recombination into a CH$_4$ molecule. Rather, this outcome corresponds to photochemistry: radicals and/or photoproducts are created. The third is then (iii) the energy is relaxed through various pathways. The weight of this third channel can be deduced from the quantification of the initial step and the first two outcomes. The absorption cross-section of pure solid CH$_4$ at Lyman-$\alpha$ (10.2 eV) has been measured by \citet{cruz-diaz_vacuum-uv_2014-1} as $1.5 \times 10^{-17}$ cm$^2$. Assuming that desorption occurs in the first 3 monolayers of the ice, we find a photodesorption yield of \textasciitilde0.05 molecules desorbed per absorbed photons at Lyman-$\alpha$. This value characterizes channel (i). \citet{oberg_effect_2010} evaluated the effective photodestruction cross-section of pure CH$_4$ ice irradiated by a Lyman-$\alpha$  dominated discharge lamp as $5 \times 10^{-19}$ cm$^2$, which corresponds to \textasciitilde0.033 destroyed molecules per absorbed photons. We can consider this value to represent channel (ii) if we make the hypothesis that photodestruction is similar at the surface and in the bulk. Both values derived for channels (i) and (ii) are low and suggest that channel (iii) dominates: most of the absorbed energy is relaxed away.

These values can be compared with the photodesorption yields per absorbed photon derived by \citet{cruz-diaz_vacuum-uv_2014} for CO, which are above unity for all wavelengths, meaning that each photon absorption leads on average to one or even multiple desorption events. The energetics in these two cases are therefore very different. The photodesorption yield per absorbed photon of CH$_4$ is closer to the one found in \citet{cruz-diaz_vacuum-uv_2014-1} for O$_2$, another molecule where photodesorption has been linked to photodissociation. 
 
\subsection{Desorption induced by CO molecules}

The results obtained on layered CH$_4$:CO ices provide some insights on CO-induced desorption. In a previous study, N$_2$ overlayer desorption induced by the underlying CO ice excitation has been shown to be a very efficient process \citep{bertin_uv_2012}. Indeed, the photodesorption yield of N$_2$ can reach up to $\sim$ 2.5 $\times$ 10$^{-2}$ molecule/photon at a photon energy of 8.3~eV when it is adsorbed onto CO ice, whereas it is almost zero when N$_2$ is pure. As can be seen in Fig. 2, the CH$_4$ desorption induced by the CO excitation barely reaches 2 $\times$ 10$^{-3}$ molecule/photon at the same energy. It is interesting, therefore, to discuss this difference of one order of magnitude as compared with the case of N$_2$. The efficiency of the indirect process for different molecules can be driven by several parameters corresponding to the different steps of the mechanism: the inter-molecular energy transfer, the intra or inter-molecular energy relaxation once it has been transferred, and the binding energy of the molecule. CH$_4$ is as volatile as CO and N$_2$, and although no data is available on the CH$_4$-CO binding energy to our knowledge, it is reasonable to assume that it would not be very different from the CH$_4$ binding energy with itself (\textasciitilde100 meV,  \citet{smith_desorption_2016}). If the energy transfer is simply a kinetic momentum transfer, since CH$_4$ is lighter than N$_2$ or CO, we would expect the transfer to be even more efficient. However, a good case can be made for the limiting step being intra or inter-molecular energy relaxation. Aside from a higher density of vibrational states into which energy can be partitioned, vibrational relaxation of polyatomic molecules is usually much faster than for diatomic molecules, due to the availability of relaxation pathways involving multiple vibration modes that do not exist for diatomic molecules \citep{nitzan_chemical_2006}. Such an explanation also fits with the observed fact that in CH$_3$OH:CO ice mixtures, CO-induced desorption of CH$_3$OH is not observed \citep{bertin_uv_2016}. Further investigations will aim at determining precisely what parameter(s) drive the efficiency of this process. 

\section{Astrophysical implications}

   \begin{table}
      \caption[]{$^{13}$CH$_4$ photodesorption yields per incident photon for pure CH$_4$ ice in various interstellar environments}
         \label{table1}
         \begin{tabular}{c c c c}
            \hline
              &  Dense cores$^a$ & Protoplanetary & ISRF$^c$ \\
              &                  & Disks$^b$      &           \\    
              & \multicolumn{3}{c}{(x 10$^{-3}$ molecules/photon)} \\
            \hline
            Pure CH$_4$ & $2.2 \pm 1.1$ & $2.2 \pm 1.1 $ & $2.0 \pm 1.0$ \\
            \hline
         \end{tabular}
         \tablefoot{UV fields are taken from $^a$\citet{gredel_c/co_1987}, $^b$\citet{johns-krull_how_2007} and $^c$\citet{mathis_interstellar_1983}}
         \label{table_yields}
   \end{table}

\subsection{Photodesorption yields of CH$_4$ from interstellar icy grains} 

By multiplying the photodesorption spectra we obtained with typical UV fields (with their integral normalized to unity) of different regions of the interstellar medium, we can derive average photodesorption yields for each of these regions. Table~\ref{table_yields} gives such values for the dense core UV field calculated by \citet{gredel_c/co_1987}, the protoplanetary disk field calculated by \citet{johns-krull_how_2007} and the interstellar radiation field (ISRF) estimated by \citet{mathis_interstellar_1983}. The UV field in dense cores and the regions of protoplanetary disks far away from the star comes from secondary UV photons generated by cosmic rays, and is dominated by Lyman-$\alpha$ photons (at 10.2~eV). The photodesorption of CH$_4$ being relatively efficient at Lyman-$\alpha$ (Fig.~\ref{fig1}), the average desorption yield for those environments is equal to the desorption yield at Lyman-$\alpha$ (Table~\ref{table_yields} and Fig.~\ref{fig1}). For the ISRF, where there is a higher contribution of lower energy photons that do not efficiently photodesorb CH$_4$, the average desorption yield is slightly lower. These photodesorption yields are obtained for $^{13}$CH$_4$, but, considering the weak difference in mass, we expect about the same values for the $^{12}$CH$_4$ isotope.

We must, however, discuss the conditions of validity of these yields. They will depend on the accuracy of the UV field models for the considered region of the ISM, but also on how well our experimental conditions can be extrapolated to astrophysical ice mantles. They are thus only entirely valid for pure, thick ices of methane deposited at 10~K, that have not been further processed. 
"Thick" in the context of photodesorption experiments can be taken as a few tens of monolayers. Photodesorption is a surface process that mostly involves the first few monolayers of the ice, and although deeper layers can play a role to some extent, 20 monolayers can thus be considered "thick". The data obtained for 1 ML and 3 ML of CH$_4$ deposited on CO ice suggest, however, a thickness dependence of the photodesorption yields for thinner layers. Another factor that can affect photodesorption is ice morphology. Ice morphology is also linked to the deposition temperature of the ice in experiments and post-processing by UV photons or cosmic ray analogs, as has been extensively studied in the case of H$_2$O \citep[e.g.][and references therein]{dartois_swift_2013}. A recent study by \citet{munoz_caro_photodesorption_2016} has highlighted the influence of the physical properties of the CO ice on photodesorption: desorption yields for CO ices deposited between 7 and 20~K vary by up to a factor of 4. This effect has been attributed to the disorder of CO molecular dipoles affecting the energy transfer between molecules, and thus hindering the indirect desorption process mentioned in section 3.2. 

Another important aspect to take into account is the molecular environment: CH$_4$ is not found pure in ice mantles. Observations \citep{boogert_observations_2015} and models \citep{garrod_formation_2011} currently point to a two-phase model of interstellar ices: one H$_2$O-rich phase, formed when hydrogenation of atoms is a dominant process and containing NH$_3$, CH$_4$ and CO$_2$, and one CO-rich phase containing CH$_3$OH and CO$_2$. According to this model, an astrophysically realistic situation would therefore rather be CH$_4$ in a water matrix, containing other elements such as NH$_3$ and CO$_2$. In a first approximation, the yields obtained for pure ices can be extrapolated to an astrophysical ice by simply weighing each yield by the relative abundance of the species. However, a different environment means different binding energies, relaxation dynamics and chemistry, all of which can significantly affect photodesorption - not to mention indirect desorption induced by different molecules. The latter is evidenced by our studies of desorption induced by CO. A more realistic case that still needs to be studied is the photodesorption of methane from an H$_2$O-rich ice. The relative parts of desorption, destruction and relaxation are likely to change in a water matrix, although it is hard to predict in which way desorption will be affected. Experiments by \citet{oberg_effect_2010} show a much increased photodestruction cross-section for CH$_4$ mixed with H$_2$O, while experiments on CO deposited on amorphous solid water (ASW) showed a decreased photodesorption, attributed to the ability of dangling OH from the water matrix to evacuate energy into the bulk very efficiently \citep{bertin_uv_2012}. A recent study by \citet{martin-domenech_study_2016} on the photodesorption of CH$_4$:H$_2$O ices only showed the desorption of reaction products (CO and H$_2$CO) initiated by the methane photodissociation. In this case, the desorption yield of intact CH$_4$ from a H$_2$O-rich ice has not been measured. More experiments, which await further beamtime, will be required to investigate all these considerations and to quantify all these effects on the CH$_4$ photodesorption yields from a water-rich or a CO-rich environment.

\subsection{Solar system}

As mentioned in section 1, photodesorption may also be relevant in astrophysical contexts other than the ice mantles of interstellar grains, for example in the solar system. CH$_4$ is classified among the supervolatile molecules: in the ultra-high vacuum conditions of our experimental set-up, its sublimation temperature is around 30K. Therefore, non-thermal desorption will likely only matter for the coldest regions of the solar system, in places where the ice is never exposed to solar light and photodesorption occurs because of the interstellar UV photons and the solar UV photons scattered by the interplanetary medium. For example, a scenario was recently published \citep{grundy_formation_2016} where the red color of the poles of Charon observed by New Horizons are suggested to originate from products of the photolysis of condensed CH$_4$ by the back-scattered solar UV photons. The CH$_4$ in this scenario comes from the evaporation of Pluto and is continuously deposited on Charon and trapped at the poles where the temperature is low enough for its condensation. This is a typical case where thermal desorption does not occur and non-thermal desorption pathways may thus play a role in the gas/ice dynamics. While in this precise case, the deposition rate (\textasciitilde $2 \times 10^8$ molecules.cm$^{-2}$.s$^{-1}$) is much higher than the photodesorption rate expected considering the measured VUV flux ($3.5 \times 10^7$ photons.cm$^{-2}$.s$^{-1}$ with mostly Lyman-$\alpha$ photons would mean \textasciitilde $7.7 \times 10^4$ desorbed molecules.cm$^{-2}$.s$^{-1}$), this scenario, as suggested by its authors, may be extended to other, smaller objects around Pluto. The example of Nix is taken, where the deposition rate is estimated to be 20 000 times lower than for Charon, in which case photodesorption would probably start being significant. 

\section{Conclusions}

Pure amorphous CH$_4$ ices at 10~K were irradiated with synchrotron light (91-177~nm -- 7-13.6~eV), and spectrally-resolved absolute CH$_4$ photodesorption yields were measured. These yields strongly depend on the wavelength, with no desorption below $\approx$136~nm. Photodesorption yields were derived for pure CH$_4$ ices for various astrophysical environments: for dense cores and disks they were found to be $2.2 \pm 1.1 \times 10^{-3}$ molecules per incident photon, equal to the value at Lyman-$\alpha$ energy, and for diffuse ISM conditions a value of $2.0 \pm 1.0 \times 10^{-3}$ molecules per incident photon was found. These yields can be added to models, keeping in mind the limits of their validity as has been discussed. Thin layered ices were also studied (1~ML of CH$_4$ above CO ice), and gave a different CH$_4$ photodesorption spectrum, with spectral signatures from both CH$_4$ and CO. This showed that both direct (i.e. the molecule absorbing a VUV photon is the one desorbing) and indirect (i.e. the molecule absorbing a VUV photon is not necessarily the one desorbing) electronic induced desorption mechanisms are at play. The relatively low efficiency of CO-induced indirect desorption in the case of CH$_4$ allowed to discuss the parameters driving this process.

\begin{acknowledgements}
We acknowledge SOLEIL for provision of synchrotron radiation facilities under the project 20150760 and we would like to thank Laurent Nahon for assistance on the beamline DESIRS. This work was supported by the French program “Physique et Chimie du Milieu Interstellaire” (PCMI) funded by the Conseil National de
la Recherche Scientifique (CNRS) and Centre National d’Etudes Spatiales (CNES). Financial support from the LabEx MiChem, part of the French state funds managed by the ANR within the investissements d'avenir program under reference ANR-11-10EX-0004-02 is acknowledged. Fundings by the Ile-de-France region DIM ACAV program is gratefully acknowledged. 
\end{acknowledgements}

\bibliographystyle{aa}
\bibliography{CH4_Biblio}

\end{document}